\documentclass[conference,10pt]{IEEEtran}
\IEEEoverridecommandlockouts

\usepackage{amsmath, amsfonts} % amssymb,
\usepackage{algorithmic}
\usepackage{graphicx}
\usepackage{textcomp}
\usepackage{xcolor}
\usepackage{subfig,paralist}
\usepackage{hyperref}
\usepackage{booktabs}
\usepackage{tabularx}
\usepackage{listings}
\usepackage{tcolorbox}
\usepackage{enumitem}
\usepackage{todonotes}

\lstdefinestyle{mystyle}{
    commentstyle=\color{codegreen},
    keywordstyle=\color{magenta},
    numberstyle=\tiny\color{codegray},
    stringstyle=\color{codepurple},
    basicstyle=\tiny\ttfamily,
    breakatwhitespace=false,
    breaklines=true,
    captionpos=b,
    keepspaces=true,
    numbers=left,
    numbersep=5pt,
    showspaces=false,
    showstringspaces=false,
    showtabs=false,
    tabsize=2,
    columns=fixed
}
\definecolor{codegray}{rgb}{0.5,0.5,0.5}
\definecolor{codegreen}{rgb}{0,0.6,0}
\definecolor{codepurple}{rgb}{0.58,0,0.82}

\newcommand{\result}[1]{%
\begin{tcolorbox}[colframe=black,boxrule=0.5pt,arc=4pt,
      left=6pt,right=6pt,top=6pt,bottom=6pt,boxsep=0pt,width=\columnwidth]%
      {\emph{#1}}
\end{tcolorbox}%
}

\AtBeginDocument{%
  }

\begin{document}

\title{Top Score on the Wrong Exam: On Benchmarking in Machine Learning for Vulnerability Detection}

\author{
\IEEEauthorblockN{Niklas Risse}
\IEEEauthorblockA{MPI-SP\\Bochum, Germany}
\and
\IEEEauthorblockN{Jing Liu}
\IEEEauthorblockA{MPI-SP\\Bochum, Germany}
\and
\IEEEauthorblockN{Marcel Böhme}
\IEEEauthorblockA{MPI-SP\\Bochum, Germany}
}

\maketitle

\begin{abstract}
According to our survey of machine learning for vulnerability detection (ML4VD), 9 in every 10 papers published in the past five years define ML4VD as a function-level binary classification problem:
\begin{center}
\emph{Given a function, does it contain a security flaw?}
\end{center}
From our experience as security researchers, faced with deciding whether a given function makes the program vulnerable to attacks, we would often first want to understand the context in which this function is called.

In this paper, we study how often this decision can really be made without further context and study both vulnerable and non-vulnerable functions in the most popular ML4VD datasets. We call a function ``\emph{vulnerable}'' if it was involved in a patch of an actual security flaw and confirmed to cause the program's vulnerability. It is ``\emph{non-vulnerable}'' otherwise. We find that in almost all cases this decision \emph{cannot} be made without further context. Vulnerable functions are often vulnerable only \emph{because} a corresponding vulnerability-inducing calling context exists while non-vulnerable functions would often be vulnerable \emph{if} a corresponding context existed.

But why do ML4VD techniques achieve high scores even though there is demonstrably not enough information in these samples? Spurious correlations: We find that high scores can be achieved even when only word counts are available. This shows that these datasets can be exploited to achieve high scores without actually detecting any security vulnerabilities.

We conclude that the prevailing problem statement of ML4VD is ill-defined and call into question the internal validity of this growing body of work. Constructively, we call for more effective benchmarking methodologies to evaluate the true capabilities of ML4VD, propose alternative problem statements, and examine broader implications for the evaluation of machine learning and programming analysis research.

\end{abstract}

\section{Introduction}
\label{sec:introduction}
In recent years, the number of papers published on the topic of machine learning for vulnerability detection (ML4VD) has dramatically increased. Because of this rise in popularity, the validity and soundness of the underlying methodologies and datasets becomes increasingly important.
So then, how exactly is the problem of ML4VD defined and thus evaluated?

In our survey of all ML4VD papers (81) published in the top Software Engineering and Security conferences and journals between January 2020 and December 2024,\footnote{Software Engineering conferences: ICSE, FSE, ASE, and ISSTA; Security conferences: S\&P, USENIX, NDSS, and CCS; Software Engineering and Security journals: TSE, TOSEM, TDSC, and TIFS.} we find that the great majority (88\%) of state-of-the-art ML4VD techniques define ML4VD as a \emph{function-level binary classification problem}: Given only the function $f$,
decide whether $f$ contains a security vulnerability or not. The technique with the lowest classification error on the evaluation dataset
is considered the best at detecting security vulnerabilities.

\begin{figure}[t]
\centering
    \begin{minipage}{\linewidth}
        \lstset{belowskip=0pt}
\begin{lstlisting}[lastline=6, frame=t, language=C, linewidth=0.95\linewidth]
TfLiteStatus ResizeOutputTensors(TfLiteContext* context, TfLiteNode* node,
                               const TfLiteTensor* axis,
                               const TfLiteTensor* input, int num_splits) {
  int axis_value = GetTensorData<int>(axis)[0];
  // [...]
  const int input_size = SizeOfDimension(input, axis_value);
\end{lstlisting}
        \lstset{aboveskip=0pt}
\begin{lstlisting}[firstline=7,lastline=7, firstnumber=7, language=C, backgroundcolor=\color{orange!30}, linewidth=0.95\linewidth]
  TF_LITE_ENSURE_MSG(context, input_size % num_splits == 0,
\end{lstlisting}
        \lstset{aboveskip=0pt,belowskip=5pt}
\begin{lstlisting}[firstline=8, firstnumber=8, frame=b, language=C, linewidth=0.95\linewidth]
                     "Not an even split");
  const int slice_size = input_size / num_splits;
  for (int i = 0; i < NumOutputs(node); ++i) {
    TfLiteIntArray* output_dims = TfLiteIntArrayCopy(input->dims);
    output_dims->data[axis_value] = slice_size;
    // [...]
    TF_LITE_ENSURE_STATUS(context->ResizeTensor(context, output, output_dims));
  }
  return kTfLiteOk;
}
\end{lstlisting}
    \end{minipage}

\caption{Context-dependent vulnerability (CVE-2021-29599) in DiverseVul dataset. If the function is called with \texttt{num\_splits=0}, it crashes with a division-by-zero in Line~7.}
    \label{fig:dependence_example}
\end{figure}

However, based on our experience, we hypothesized that it might not always be possible to determine whether a function is vulnerable or not without additional context. We call these vulnerabilities \emph{context-dependent}. Consider the example in \autoref{fig:dependence_example}. If this function from the DiverseVul benchmark dataset \cite{diversevul} is called with \texttt{num\_splits} set to zero, it will crash with a division-by-zero in Line~7.
However, without knowing whether this function can ever be called with \texttt{num\_splits} set to zero, we cannot reliably decide if the division-by-zero could actually be observed. This function parameter might as well be properly validated by every caller function. This context-dependency problem is well-studied for other approaches to vulnerability detection, such as \emph{static analysis} \cite{manifest_vs_latent, phasasr} (where it is handled, e.g., by defining suitable preconditions) or \emph{software testing} \cite{desikan2006software} (which distinguishes, e.g.,  between system- and unit-level testing).

In this paper, we set out to quantify the prevalence of context-dependent vulnerabilities in the most popular datasets and end up revealing a fundamental flaw in the most-widely used evaluation methodology that is underpinning the progress of the nascent research area of ML4VD. We find that the vulnerability of a function cannot be decided without further context for more than 90\% of functions. This includes functions with both types of labels, \texttt{vulnerable} or \texttt{secure}: If the right context existed, a function labeled as \texttt{secure} would make the program vulnerable. Respectively, only \emph{because} the right context exists, a function labeled as \texttt{vulnerable} makes the program actually vulnerable.

Given our findings, we conclude that the prevailing problem statement of ML4VD as a function-level classification problem is inadequate. The reported results in the literature, which are based on this problem statement, may not accurately reflect the true capabilities of the evaluated techniques at the task of vulnerability detection. In other words, there is currently no evidence that ML4VD techniques are actually capable of identifying security vulnerabilities at the function-level.

But why do ML4VD techniques still achieve high scores at this binary classification task when there is demonstrably not enough information in over 90\% of samples (even after addressing label inaccuracies)? We identify spuriously correlated features as a potential reason. Training simple models, like a gradient boosting classifier using only word counts and disregarding code structure, we achieved results comparable to those of state-of-the-art ML4VD models. This suggests that ML4VD techniques only \emph{appear} to perform well due to the chosen evaluation methodology. During classification,
ML4VD techniques rely on spuriously correlated features to achieve high scores and do not genuinely detect vulnerabilities.

To shift the field towards more context-aware evaluation of vulnerability detection methods, we discuss potential alternative problem statements and suggest ideas for future work. We also examine the broader implications for ML4VD and other fields. By discussing these aspects we aim to foster more valid and reliable research in the area of machine learning for vulnerability detection.

\vspace{0.1cm}
\noindent
In summary, this paper makes the following contributions:
\begin{itemize}
    \item We analyze all papers published at the top Software Engineering conferences, Security conferences, and journals over the last five years and find that the great majority (88\%) of state-of-the-art ML4VD techniques define ML4VD as a function-level classification problem.
    \item We reveal a fundamental flaw of the function-level classification problem: The vulnerability of a function cannot be decided without further context for more than 90\% of functions in the top-most widely-used datasets.
    \item Why do ML4VD techniques still achieve high scores at the function-level classification problem? We demonstrate that they may rely on spurious features to achieve high scores without genuinely detecting vulnerabilities.
\end{itemize}

\section{Background}
\label{sec:related_work}

The context dependency problem---i.e., the determination whether a function causes a program to be vulnerable depends on the context that is \emph{external} to the given function---has been well-studied in static analysis and software testing.

\textbf{Static Analysis.} We distinguish between inter-procedural and intra-procedural analysis, where the former is concerned with the analysis of the entire system and the latter with the analysis of individual functions.
Intra-procedural static analysis tools often struggle with false positives due to context-dependency. This is a core problem in static analysis, which has received much attention in the literature \cite{nielson_book}. In practice, the context-dependency check is ultimately delegated to the user of the tool (which, however, similarly complicates the experimental evaluation in the absence of a user). A programmatic approach is to (manually or automatically) annotate the function with a precondition (e.g., in the style of Hoare logic \cite{hoare_logic}) that encodes assumptions about the \emph{valid} state space at the start of the function call. Function summaries or type systems provide other mechanisms to encode the valid context under which the function is (meant to be) called. Despite these efforts, discerning the specific conditions under which a bug manifests remains a challenge. For example, a static analyzer might flag potential bugs based on the analysis of code patterns but cannot always discern the specific conditions under which a bug manifests \cite{sast1,sast2}. Hence, Le et al. \cite{manifest_vs_latent} propose to differentiate between manifest bugs, which are context-independent, and latent bugs, which depend on preconditions in the calling context. Manifest bugs can be reported without making any assumptions about the calling context, whereas latent bugs cannot.

\textbf{Software Testing.} We distinguish between system- and unit-level testing \cite{challenges}, where the former is concerned with testing the entire system  \cite{FSE25-mendelfuzz,chatafl} and the latter with testing individual system units, like functions \cite{ICSE25-invivo,entropic2}.
Automated unit test generation often falls short of identifying issues that only occur under specific conditions or in particular environments. As noted by Harrold and Orso \cite{harrold_unit_testing}, unit tests can produce false positives when they are not adequately designed to account for the broader context in which a function operates. This can lead to an inflated number of reported issues that are not actual bugs, thereby complicating the debugging process. Hence, property-based testing \cite{claessen2000quickcheck,jqf,zest} requires users to define function preconditions in addition to assertions to ensure that the assertions fire only under \emph{valid} function parameters (i.e., under valid context).

To summarize, both the static analysis and software testing research communities acknowledge the problem of context-dependency and address it through various techniques aimed at reducing the number of false positives.

\textbf{Machine Learning (ML4VD)}. Several colleagues have previously raised concerns about the validity of this growing body of work.
Ullah et al. \cite{s_p_sec_holmes} study the performance of LLMs for vulnerability detection and find that results are nondeterminstic, the reasoning is incorrect and unfaithful, and that they perform poorly in real-world scenarios.
Risse and B{\"o}hme \cite{risse2023limits} find that the capability of state-of-the-art ML4VD techniques to distinguish between buggy and patched function is barely better than a coin flip.
Croft et al. \cite{data_quality_for_se_datasets} study the quality of benchmark datasets used in the evaluation of ML4VD techniques. They find that a large percentage of vulnerability labels in real-world datasets are in fact inaccurate and many data points were even duplicated (sometimes with inconsistent labels). In this paper, we reproduce their experiment which allows us to identify and study \emph{actually} vulnerable functions in terms of their context-dependence.
However, our question remains: Can the vulnerability of a function be decided without further context in the first place? Only if it can, we need to be concerned about the quality of our datasets and evaluation procedures.

Another stream of works studies the assumption that a vulnerability is localized within a single function.
For instance, Sejfia et al. \cite{icse24_mbu} realize that the patches of some vulnerabilities span across multiple ``base units'' (e.g, functions) and measure the performance of vulnerability detectors to \emph{individually} classify \emph{all} units belonging to the same vulnerability as \texttt{vulnerable}---observing a substantial drop in performance. They also recommend to ensure that the train/validation/test split honors the grouping of units by vulnerability.
Li et al. \cite{icse_interprocedural_multi_function} study the default assumption that the vulnerability-\emph{patching} functions are labeled as \texttt{vulnerable} and find that the patch and its "trigger" (i.e., the code where the vulnerability manifests, e.g., as an array out-of-bounds write, a double-free, etc.) exist in \emph{different} functions for about a quarter of studied vulnerabilities.\footnote{To facilitate this evaluation, they also introduce a slicing-based approach, VulTrigger, to identify the trigger \emph{given the patch}.} If we labeled functions containing the vulnerability-\emph{triggering} code as \texttt{vulnerable}, function-level vulnerability detectors would turn out much less effective, they find.

However, they still define ML4VD as a unit-level classification problem. How prevalent is the function-level problem statement for ML4VD in the literature (\S\ref{sec:literature_survey})? Is it a valid problem statement to begin with (\S\ref{sec:study_design}, \S\ref{sec:results})? In answering these questions, our paper aims to uncover the shortcomings of current evaluation methodologies and emphasizes the urgent need for more context-aware approaches to accurately assess the true capabilities of ML4VD techniques in detecting security vulnerabilities. 

\textbf{Spurious features.} If we find that the decision whether a function is \texttt{vulnerable} or not cannot be made without further context that is external to that function, then how do we explain the high values of classifier performance? Maybe a classifier learns to predict well based on features that are spuriously correlated with the \texttt{vulnerable} label. Indeed, the phenomenon of spurious correlations in ML4VD is well-studied.
For instance, Arp et al. \cite{dos_and_donts} provide initial evidence that machine learning techniques can learn to predict labels correctly based on artifacts in code snippets without addressing the actual security task at hand (spurious correlation).
Risse et al. \cite{risse2023limits,risse2023src} demonstrate that ML4VD techniques overfit to label-unrelated features using semantic-preserving transformations and further that making the models robust against semantic-preserving transformation only introduces overfitting to the \emph{specific} transformations used.

Furthermore, several studies have proposed ML4VD techniques that are aimed to be less affected by spurious correlations. For example, Rahman et al. \cite{icse_24_causalvul} identify spuriously correlated variable and API names and disable their use during inference time of their model.
Cao et al. \cite{ase_24_snopy} completely remove code that is \emph{not} related to the vulnerability from the code of the \texttt{vulnerable} function during training by leveraging patch commit information. They develop a novel graph-based neural network architecture for the function-level vulnerability classification task to classify vulnerabilities based on causal features only.

While existing work studies and minimizes the impact of spurious correlations using structural representations or semantic-preserving transformations of the vulnerable code (i.e., perturbations on train and/or test data), we provide substantial evidence of this phenomenon using an even simpler and technique-independent method: A high measure of classifier performance is achieved even by a simple classifier if there are no program semantics but only word counts available as features. In this way, our work provides an orthoganal explanation of the surprising results of ML4VD even though the problem statement is ill-defined.

\section{Literature Survey}
\label{sec:literature_survey}

Based on our prior knowledge of the literature, we hypothesized that the majority of recent studies in machine learning for vulnerability detection (ML4VD) define ML4VD as a binary classification problem; given a function, determine whether the function contains a security vulnerability. To determine the prevalence of this approach in the ML4VD literature, we conducted a literature survey. Our goal was to address the following two research questions:

\begin{enumerate}
    \item \textbf{Problem Statement:} What proportion of ML4VD publications defines ML4VD as deciding whether a given function contains a vulnerability?
    \item \textbf{Datasets:} Which datasets do they use to evaluate their techniques empirically?
\end{enumerate}

\subsection{Methodology}

To ensure a thorough and unbiased analysis of the recent literature on machine learning for vulnerability detection (ML4VD), we adopted a systematic methodology based on established guidelines for conducting literature reviews in Software Engineering research \cite{Kitchenham2007Guidelines, Wohlin2014Guidelines}. Our approach encompassed three primary phases: defining the scope, selecting relevant papers, and systematically analyzing the selected papers.

\begin{figure*}[t]
    \subfloat[ML4VD papers per problem statement granularity.\label{fig:survey:a}]{%
        \includegraphics[width=0.48\linewidth]{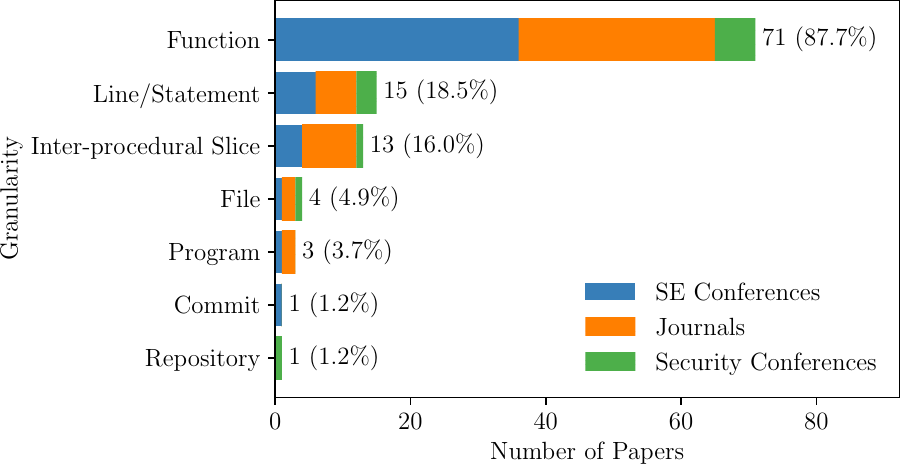}
    }
    \hfill
    \subfloat[ML4VD papers per year.\label{fig:survey:b}]{%
        \includegraphics[width=0.48\linewidth]{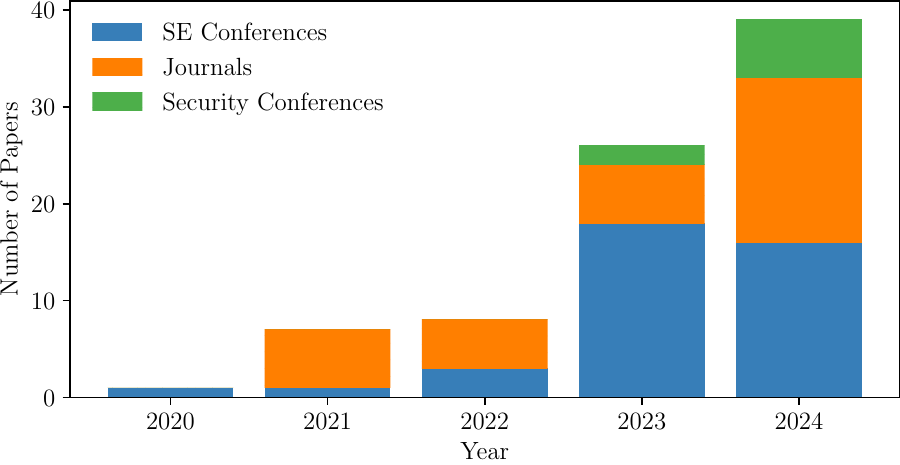}
    }
    \caption{Literature survey results for the 81 ML4VD papers we identified in the top Software Engineering (SE) and Security conferences and journals. \autoref{fig:survey:a} shows how the papers define the problem of ML4VD. Note that a paper may use multiple granularities, which explains why the numbers in \autoref{fig:survey:a} do not add up to 100\%. \autoref{fig:survey:b} shows how many papers were published each year since 2020.}
    \label{fig:survey}
\end{figure*}

\textbf{Scope.} Our literature survey focuses on the past five years, targeting publications from 2020 to 2024. Given the prominence and impact of certain venues in the fields of Software Engineering and Security, we selected a set of the most impactful conferences and journals. For Software Engineering, we included International Conference on Software Engineering (ICSE), Foundations of Software Engineering (FSE), International Symposium on Software Testing and Analysis (ISSTA), and Automated Software Engineering (ASE). For Security, we incorporated Symposium on Security and Privacy (S\&P), Network and Distributed System Security Symposium (NDSS), Computer and Communications Security (CCS), and USENIX Security Symposium (USENIX). Additionally, we considered the top journals in these domains, including Transactions on Software Engineering and Methodology (TOSEM), Transactions on Software Engineering (TSE), Transactions on Information Forensics and Security (TIFS), and Transactions on Dependable and Secure Computing (TDSC). These venues were chosen for their reputation in publishing high-quality and influential research in the Software Engineering and Security fields (CORE\footnote{CORE: \url{https://www.core.edu.au/home}} rank A* or A).

\textbf{Criteria for Paper Selection.} To identify relevant papers, we established clear selection criteria. The papers must be full-length papers published in the proceedings of the listed in-scope conferences or journals between 2020 and 2024. We do not include short papers (e.g. for PhD student competitions), since they usually represent work-in-progress. We also do not include pure literature reviews if they do not present new experimental data. The papers must fit the topic of machine learning for vulnerability detection (ML4VD), which means that they either propose a new ML-based technique for vulnerability detection, or that they present experiments which use existing techniques. We include all types of machine learning techniques, e.g. graph neural networks, large language models, or convolutional neural networks. We explicitly focus on vulnerability detection for software, which excludes related fields, e.g. hardware or blockchain vulnerability detection. To facilitate the paper identification process, we defined a set of keywords that are indicative of the research focus in this area: vulnerability, vulnerable, detect, detection, discovery, machine, learning, artificial, intelligence, AI, ML, deep, graph, neural, network, large, language, model. We scanned the titles and abstracts of all in-scope papers using multiple tools: IEEE Xplore Advanced Search\footnote{IEEE Xplore Advanced Search: \url{https://ieeexplore.ieee.org/search/advanced}} for the IEEE venues (ICSE, ASE, S\&P, TSE, TDSC, and TIFS), ACM DL Advanced Search\footnote{ACM DL Advanced Search: \url{https://dl.acm.org/search/advanced}} for the ACM venues (ICSE, FSE, ASE, ISSTA, CCS, and TOSEM) and Google Scholar Advanced Search \footnote{Google Scholar Advanced Search: \url{https://scholar.google.com/}} for the independent venues (NDSS and USENIX). This resulted in more than 2500 potential papers, which were flagged for further review.

\textbf{Filtering Process.} As a first step, we manually filtered the flagged titles to specifically exclude those that most likely do not fit the topic of ML4VD. One of the co-authors of this paper, a Software Security researcher with publications at the top Software Engineering and Security conferences on the topic of ML4VD, manually checked each of the paper titles to facilitate the first round of filtering. This resulted in a preliminary list of 109 papers, which we include as supplementary material to this paper. The author then reviewed each of the 109 abstracts and, where necessary, the full texts to assess their relevance based on our selection criteria. This resulted in a final list of 81 papers matching the selection criteria. The 28 excluded papers are either off-topic (e.g. vulnerablity management) or do not leverage ML-based techniques (e.g. static analysis).

To answer research questions 1) and 2), we documented how ML4VD is defined, and the datasets used for empirical evaluation within each of the 81 identified ML4VD papers.

\subsection{Results}

Our literature survey identified 81 papers that met our selection criteria and were included in our analysis. The full list of papers is part of the supplementary material of this paper. 39 papers were published in the Software Engineering conferences, 8 in the Security conferences, and 34 in the journals. The distribution of these papers across the years and conferences, as displayed in \autoref{fig:survey:b}, indicates a clear trend: the number of papers focused on ML4VD is increasing annually. This upward trajectory underscores the growing interest and significance of this research area within the Software Engineering and Security communities. Additionally, we found a large variety of machine learning techniques being employed, utilizing different data representations (graph-based, token-based), model architectures (e.g., large language models, graph neural networks, convolutional neural networks), and learning algorithms (e.g., contrastive learning).

\textbf{Problem Statement.} \autoref{fig:survey:a} shows how the 81 papers included in our literature survey define the problem of ML4VD. To our surprise, the great majority (88\%) of papers across the in-scope Software Engineering conferences (92\%), Security conferences (75\%) and journals (85\%) define ML4VD as a binary function-level classification problem. This problem statement, which focuses on determining whether a given function contains a security vulnerability, has become the dominant problem statement in the field. The yearly trend seems to be increasing since 2021, with 95\% of papers relying on a function-level problem statement in 2024 (2020: 100\%, 2021: 57\%, 2022: 75\%, 2023: 88\%, 2024: 95\%). 15 out of 81 (19\%) papers define ML4VD at the line-/statement-level and 13 out of 81 papers (16\%) at the inter-procedural slice-level. A inter-procedural slice is a subset of a program's statements affecting a particular computation, spanning across multiple functions or procedures. Notably, the two most popular problem statements (function-level and line-/statement-level) exclusively rely on information within a single function for classification (intra-procedural). The remaining problem statements (file-, program-, commit-, and repository-level) all rely on inter-procedural information, but are only used by a minority of papers (10 out of 81). 

\begin{figure}[t]
    \includegraphics[width=\linewidth]{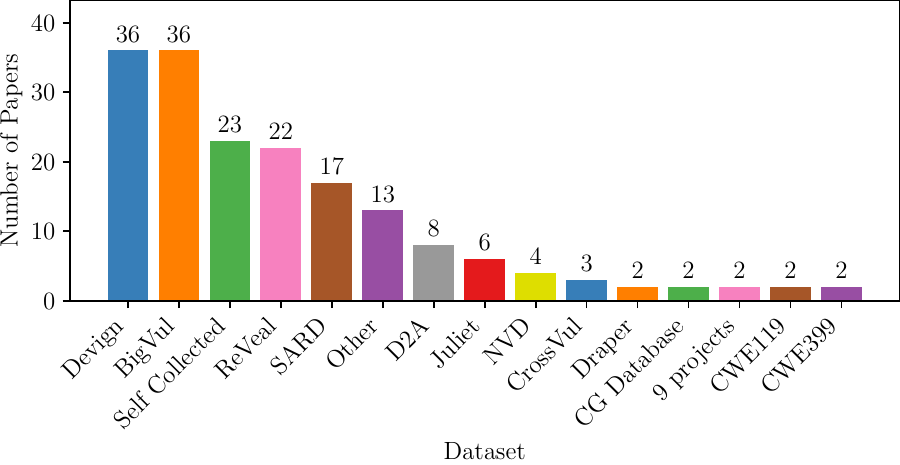}
    \caption{Datasets that are used by ML4VD papers published at the top Software Engineering and Security conferences and journals over the last five years. Datasets that were used only once are displayed as "Other".}
    \label{fig:datasets}
\end{figure}

\textbf{Datasets.} Regarding datasets, the majority of the papers relied on a limited set of popular datasets for empirical evaluation. \autoref{fig:datasets} shows all datasets that were used by the 81 papers and the number of times they were used. Specifically, BigVul \cite{bigvul} and Devign \cite{devign} were both used by 36 papers, and ReVeal \cite{reveal} by 22 papers. Notably, 53 out of the 81 papers utilized at least one of these three datasets, reflecting their dominance in the field.

\result{\textbf{Literature Survey.} The great majority of ML4VD papers (88\%) published at top Software Engineering conferences (ICSE, FSE, ISSTA, and ASE), Security conferences (S\&P, USENIX, NDSS, and CCS), and journals (TSE, TOSEM, TIFS, and TDSC) over the past five years define ML4VD as a binary classification problem: Given an isolated function $f$, decide whether $f$ contains a security vulnerability. Moreover, three datasets—BigVul, Devign, and ReVeal—are overwhelmingly favored, with 53 out of the 81 papers using at least one of these datasets for empirical evaluation.}

\section{Empirical Study Design}
\label{sec:study_design}

We study how prevalent context-dependent vulnerabilities are in the top-most widely-used benchmarks for ML4VD and evaluate the degree to which we can correctly classify even when the vulnerable code is \emph{hidden}, i.e., even when only coarse feature values such as word counts are available to a simple classifier.
More generally, we are interested in the threats to validity of the benchmarking methodology that is used most-widely in ML4VD research: Are empirical claims from benchmarking results about the real-world performance of ML4VD techniques actually reliable?

\vspace{0.1cm}
\noindent
Specifically, we ask the following research questions:
\begin{itemize}[leftmargin=0.3cm]
\item \textbf{RQ.1 Can we cast vulnerability detection as a function-level binary classification problem in ML4VD?}
\begin{enumerate}[label=(\alph*),leftmargin=0.5cm]
    \item \textbf{Noisy Labels.} \emph{What proportion of functions labeled as vulnerable actually contain security vulnerabilities?} Before we can study the prevalence of context-dependency, we first address the noisy-label problem \cite{data_quality_for_se_datasets} and identify those functions that are actually vulnerable.
    \item \textbf{Context-dependent Vulnerability.} \emph{What proportion of vulnerable functions would \emph{not} be vulnerable if the appropriate external context did \emph{not} exist?}
    Given a function that is actually vulnerable (because it is later fixed to remove a vulnerability), how often can we decide vulnerability based on the function's code alone?
    \item \textbf{Context-dependent Security.} \emph{What proportion of non-vulnerable functions could be vulnerable if an appropriate external context would exist?}
    Given a function that is \emph{not} vulnerable within the context of this program, how often can we find a setting in which this function would be considered to contain a vulnerability?
\end{enumerate}
\item \textbf{RQ.2 Can we achieve a high classification performance even when the root cause of the vulnerability is hidden?} How can we explain the excellent performance of ML4VD on these widely-used benchmarks despite this severe flaw in the problem statement? Do the popular datasets contain properties that ML4VD techniques can exploit to achieve high scores without actually detecting vulnerabilities?
\end{itemize}

\begin{figure}[t]
    \includegraphics[width=\linewidth]{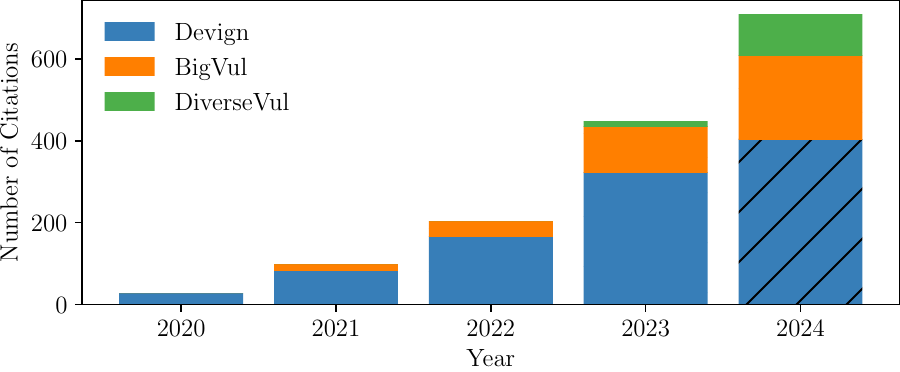}
    \caption{Popularity measured by citations of the datasets we selected for our empirical study. }
    \label{fig:citations}
\end{figure}

\subsection{Methodology RQ.1}
\label{sec:empirical_study_process}

\textbf{Selection Criteria.} In order to select datasets that represent the state-of-the-art of ML4VD benchmarks, we chose the top-most widely-used datasets from the ML4VD literature (via our literature survey) for which the authors provide links to patch commits and/or CVE websites (required for manual labeling). Additionally, we reviewed unpublished literature for emerging datasets (pre-prints from arXiv\footnote{arXiv: \url{https://arxiv.org}}).

\textbf{Selected datasets.} Based on our selection criteria, we chose the BigVul \cite{bigvul}, Devign \cite{devign}, and DiverseVul \cite{diversevul} datasets. The selection of BigVul and Devign resulted from our literature survey, as these datasets were the most popular among the 81 ML4VD papers we analyzed. However, we excluded ReVeal \cite{reveal} from our study since the authors did not publish patch commit IDs or CVEs, which are necessary for determining vulnerability and context dependence. Additionally, we included the DiverseVul dataset \cite{diversevul}, a recently published dataset (2023) that has seen significant use in many yet unpublished papers based on our review of unpublished literature. \autoref{fig:citations} illustrates the citations of the selected datasets measured by Google Scholar on January 16, 2025. All three datasets are becoming increasingly popular, with each having more than 100 citations in 2024 alone.

\textbf{Sampling.} From each of the three datasets, we randomly selected 100 samples labeled as vulnerable, using a sample size inspired by related studies on data quality, such as the one by Croft et al., which used a sample size of 70 \cite{data_quality_for_se_datasets}. We have published the reproducible script and the sampled functions as part of our GitHub repository, which is available at \url{https://github.com/niklasrisse/TopScoreWrongExam}. The resulting 300 functions come from patch commits published between 2010 and 2022, covering 80 unique open-source projects (BigVul: 25, Devign: 2, DiverseVul: 60). While we did not formally categorize the vulnerability types, we frequently observed issues such as out-of-bounds writes/reads, improper restriction of operations within memory buffer bounds, improper input validation, and use-after-free vulnerabilities.

\textbf{Labeling Process.} In a time-intensive process (more than 150 hours), two Software Security researchers (co-authors of this paper) reviewed each of the 300 functions independently. The process involved opening and understanding the corresponding patch commit on GitHub and, if available, reviewing the CVE (Common Vulnerabilities and Exposures) report for additional context.

\begin{figure*}[t]
    \centering
    \includegraphics[width=\linewidth, trim={0.6cm 0 0.6cm 0},clip]{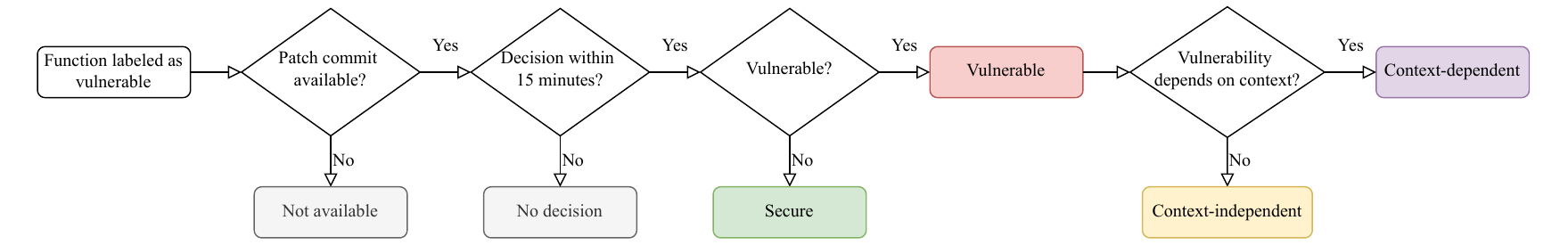}
    \caption{\textbf{Empirical Study Design:} Our manual labeling process to determine what proportion of security vulnerabilities in popular datasets can be detected without considering additional context beyond the function-level.}
    \label{fig:methodology_1}
\end{figure*}

\autoref{fig:methodology_1} visualizes the complete process that the two researchers employed for each individual function. In all cases where the patch commit was available, and a decision could be made within 15 minutes, they assigned one of the following labels:

\begin{itemize}
    \item \textbf{Secure (0):} The function was not the source of the security vulnerability, or there was no vulnerability addressed by the patch commit.
    \item \textbf{Vulnerable (1):} The function was the source of the security vulnerability.
\end{itemize}
Additionally, they added a short explanation in natural language to justify the assigned label, which we used to analyze the results more in-depth. For the first labelling step (secure or vulnerable), the labels of the two researchers agreed for 82\% of all functions, resulting in a Cohen Kappa value of 0.64 \cite{Cohen1960ACO}, which implies substantial agreement according to the guidelines provided by Landis and Koch \cite{Landis1977TheMO}. All cases of disagreement were resolved through discussion and resulted in a single label/explanation for each function. In the case that only one of the researchers was able to make a decision within 15 minutes, the researchers also entered a discussion for resolution, which resulted in a joint decision (vulnerable/secure) for all of these cases. Only if both researchers were unable to make a decision after 15 minutes, the final label was "No decision". In cases where the source of a vulnerability was ambiguous, e.g. if a vulnerability in a function could be fixed both by adding a check within the function (callee) or within a calling function (caller), the researchers only considered the callee to be vulnerable.

For functions labeled as vulnerable (indicating they actually were the source of the vulnerability), the two researchers conducted a second round of labeling to determine context dependency:

\begin{itemize}
    \item \textbf{Context-independent (0):} The vulnerability could be detected without considering any additional context beyond the function itself. These vulnerabilities are self-contained within the function’s code.
    \item \textbf{Context-dependent (1):} The security vulnerability cannot be accurately identified without considering additional context beyond the function. This category includes vulnerabilities that rely on external functions, global variables, or interactions with other parts of the codebase.
\end{itemize}

Again, the researchers provided a short explanation in natural language to justify the context label. For the second labelling step (context dependence), the labels of the two researchers agreed for 98\% of all functions, resulting in a Cohen Kappa value of 0.96 \cite{Cohen1960ACO}, which implies almost perfect agreement according to the guidelines provided by Landis and Koch \cite{Landis1977TheMO}. Again, all cases of disagreement were resolved through discussion and resulted in a single label/explanation for each function.

\textbf{Labeling Expertise.} Both Software Security researchers that carried out the labeling process have considerable experience in the Software Security research field (C/C++ security, in particular). The first researcher has three years of experience in Software Security research, published multiple papers on the topic of vulnerability detection at the top Security and Software Engineering conferences, and worked as a Software Engineer in industry for two years. The second researcher has five years of experience in Software Security, published multiple pre-prints on the topic (currently under submission at top conferences), and participated in discovering three CVEs, which are actually part of at least one of the three datasets we investigated for the empirical study.

\textbf{Reproducibility.} To ensure the reproducibility of our empirical study and to provide transparency in our research, we have made all related scripts and data publicly available. All resources can be accessed as part of our GitHub repository, which is available at \url{https://github.com/niklasrisse/TopScoreWrongExam}.

\subsection{Methodology RQ.2}

In order to test whether the three widely-used datasets (Devign, BigVul, and DiverseVul) contain spuriously correlated features that can be exploited to achieve a high performance without actually detecting security vulnerabilities, we trained a simple classifier (Gradient Boosting Classifier) to detect vulnerabilities based on word counts only, completely disregarding the structure of the code. After training, we evaluated the resulting models on the evaluation subsets of the respective datasets, again based on word counts only.

\textbf{Metrics.} We selected the f1-score as our primary performance metric, since it is the dominant metric in the literature we surveyed for \autoref{sec:literature_survey}. Additionally, we report accuracy for the Devign dataset, because it is a relatively balanced dataset (47.9\% vulnerable functions). Since both BigVul (4.5\% vulnerable functions) and DiverseVul (5.4\% vulnerable functions) are heavily imbalanced towards functions labeled as secure, using accuracy as a performance metric for model comparison could be misleading \cite{leaning_imbalanced_data}. For example, in a dataset with 95.5\% of functions labeled as secure (such as BigVul), a technique can achieve an accuracy of 95.5\% by classifying all functions as secure. This is why we chose to omit accuracy for BigVul and DiverseVul. 

\textbf{Selected Techniques.} Based on the literature we surveyed for \autoref{sec:literature_survey}, the state-of-the-art techniques for the Devign dataset are SNOPY \cite{ase_24_snopy} with 67.86\% f1-score and PDBERT \cite{icse_24_dependencies_pretraining} with 67.61\% accuracy. For the BigVul dataset, the state-of-the-art technique is DeepDFA \cite{icse_24_bigvul_deep_dfa} with 96.46\% f1-score. According to the benchmark provided by Chen et al. \cite{diversevul}, CodeT5 Small \cite{wang-etal-2021-codet5} is the best performing technique for the DiverseVul dataset with 48.28\% f1-score. 

\section{Results}
\label{sec:results}

The goal of our empirical study was to investigate whether vulnerability detection as a function-level binary classification problem is an adequate problem statement for ML4VD.

\subsection*{RQ.1-a\quad Noisy Labels}
\label{sec:results:rq_1:a}

\begin{figure*}[t]
    \subfloat[\textbf{BigVul}\label{fig:results:rq_1:a}]{%
        \includegraphics[width=0.325\linewidth]{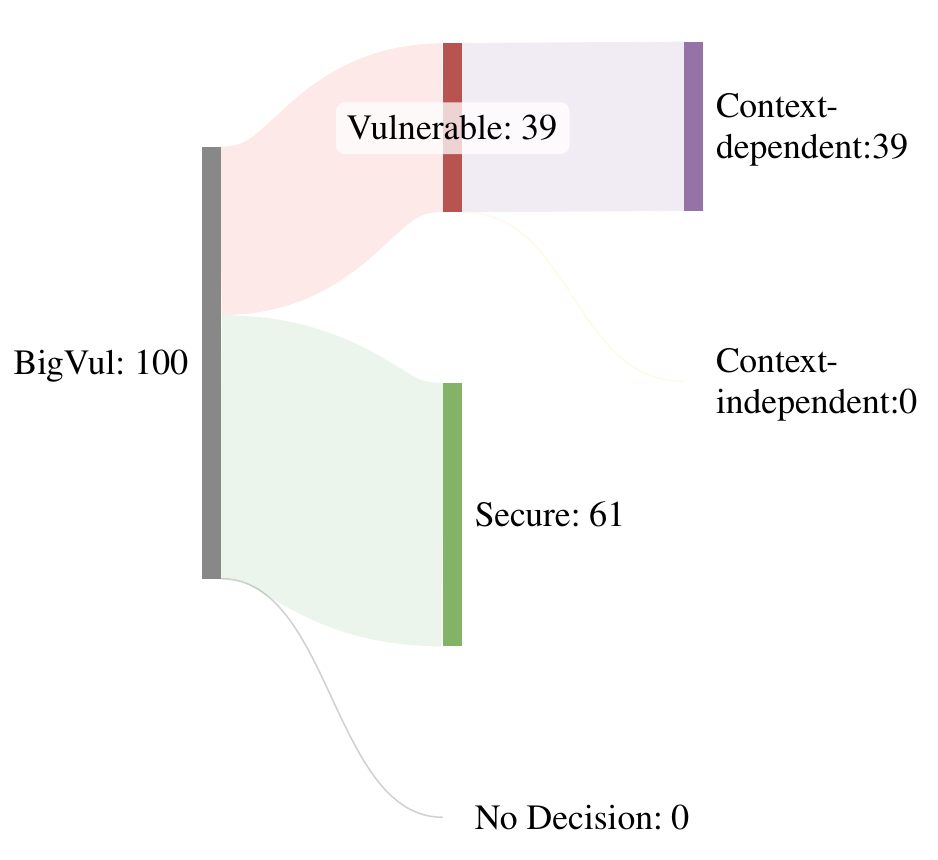}
    }
    \hfill
    \subfloat[\textbf{Devign}\label{fig:results:rq_1:b}]{%
        \includegraphics[width=0.325\linewidth]{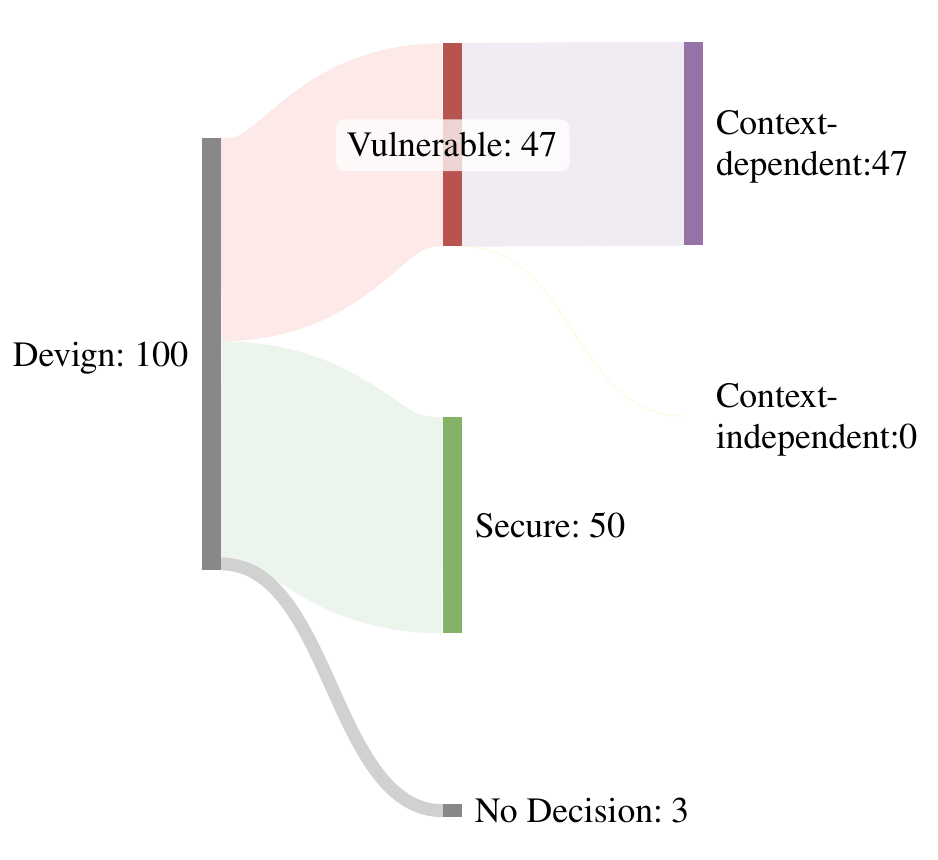}
    }
    \hfill
    \subfloat[\textbf{DiverseVul}\label{fig:results:rq_1:c}]{%
        \includegraphics[width=0.325\linewidth]{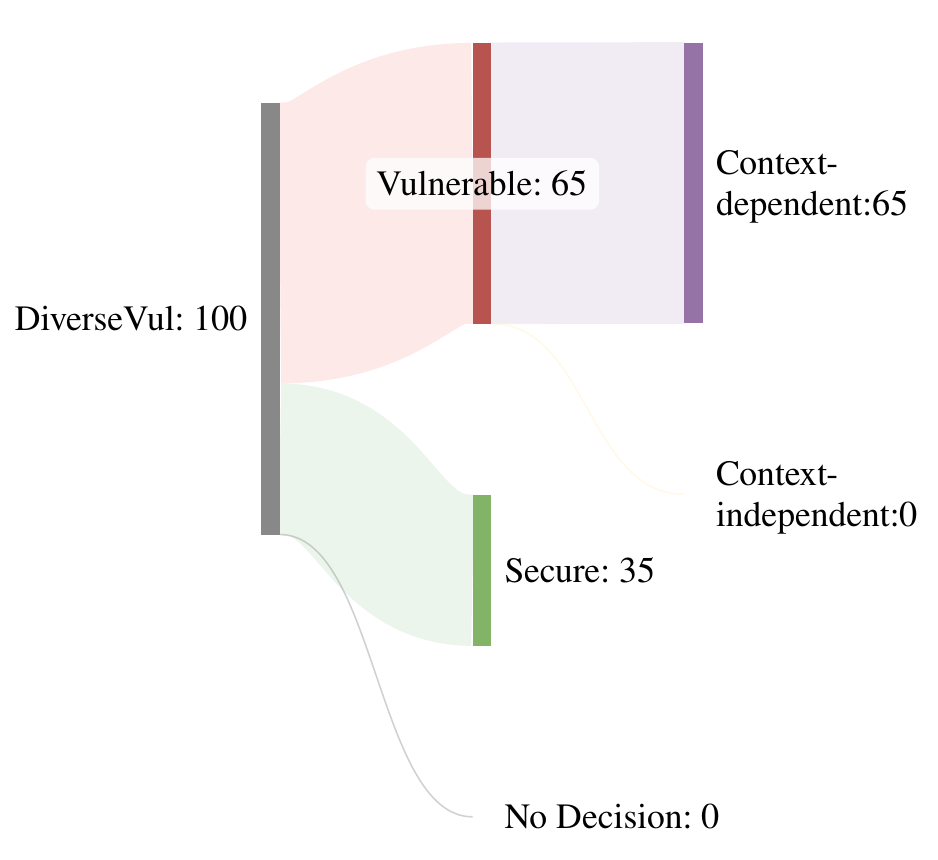}
    }
    \caption{\textbf{Empirical Study Results:} For each dataset, we manually analyzed 100 functions that were labeled as vulnerable. For the subset of functions that were actually vulnerable (red), we then determined whether the vulnerability could be detected based on the function's code alone. If yes, we marked the function as context-independent (yellow). Otherwise, as context-dependent (purple). }
    \label{fig:results:rq_1}
\end{figure*}

The first decision for each function included in our empirical study was to determine whether the function, in fact, contains a security vulnerability. This is necessary since one of the main goals of our study is to determine the proportion of \emph{actually} vulnerable functions that are dependent on additional context. From Croft et al. \cite{data_quality_for_se_datasets}, we know that for popular datasets only a subset of functions labeled as vulnerable actually contains security vulnerabilities. Specifically, they found at least 20\% of labels for the Devign dataset and 45.7\% of labels for the BigVul dataset to be inaccurate. 

\textbf{Results.} \autoref{fig:results:rq_1} shows the results of our empirical study. We were able to find the patch commit for all functions included in the empirical study, which is why the label "Not Available" was omitted from \autoref{fig:results:rq_1}. For RQ.1-a), we are interested in the proportion of functions labeled as vulnerable that actually contain security vulnerabilities, which are displayed in the left part of each subfigure (vulnerable vs. secure). Out of the 100 functions per dataset that were originally labeled as vulnerable, we found only 39\%-65\% to actually contain security vulnerabilities. 35\%-61\% do not contain security vulnerabilities and were therefore labeled to be secure. For 0\%-3\%, we were not able to make a decision after 15 minutes. 

\result{\textbf{Noisy Labels (RQ1-a).} Out of the 100 functions per dataset that were originally labeled as vulnerable, only 39\%-65\% (Devign: 47\%, BigVul: 39\%, DiverseVul: 65\%) actually contain security vulnerabilities.}

Based on our evidence, we can confirm the findings of Croft et al. \cite{data_quality_for_se_datasets}. The differences in label accuracy, especially for the Devign dataset (80\% accurate labels found by them vs. 47\% found by us), might be explained by two reasons: 
First, Croft et al. considered label accuracy for \emph{all} functions, and secure functions are more likely correctly labeled. In RQ.1-a, we only establish label accuracy for functions labeled as vulnerable.
Second, Croft et al. establish label accuracy more conservative than us, where a function can be considered correctly labeled as vulnerable if it calls a function known to be vulnerable. In this case, we decided that only the actual vulnerable function should be labeled as vulnerable because the calling function would not be vulnerable if the actual vulnerable function would be fixed.

Based on our quantitative results and the explanations we generated for each of the functions, we performed an in-depth analysis to determine potential reasons for the label inaccuracy we observed. The results are displayed in \autoref{table:reasons:label-inaccuracy}.

\textbf{Patch Commit Identification.} The first reason for label inaccuracy are errors during the process of identifying patch commits. From the original papers \cite{devign, diversevul, bigvul}, we know that all three datasets start their data collection process by identifying vulnerability-patching commits in popular open-source software repositories. However, the Devign dataset identifies these commits \emph{only} by filtering the commit messages for security-related keywords. We observe that 54\% of falsely labeled functions in our sample of the Devign dataset originate from this automatic identification process. The triangulation via other data sources (e.g. CVE databases) makes this reason occur less frequently for the other datasets (21\% for BigVul and 3\% for DiverseVul).
 
\textbf{Structural Changes.} The second reason for label inaccuracy is structural changes. All three datasets included in our study assume that all functions changed by a vulnerability-patching commit were vulnerable before the patch was applied. However, according to our results, only a subset of the functions changed by a patch commit is actually vulnerable, while other functions could be changed to address structural changes that are a consequence of fixing the actual vulnerable function. For instance, fixing a buffer overflow may require adding a buffer size parameter to the function call wherever the function is called.
This can lead to false labels if all functions changed by the patch commit are considered to be vulnerable before the patch was applied. In fact, 38\%-66\% of falsely labeled functions can be attributed to structural changes.

\begin{table}[t]
%\scriptsize
\centering
  \caption{\textbf{Reasons for Inaccurate Labels.} The different reasons for inaccurate labels we observed and their prevalence in the three datasets.}
  \label{table:reasons:label-inaccuracy}
  \begin{tabular}{l | c c c}
    \toprule
     & Patch Commit & Structural & Unrelated \\
    Dataset & Identification & Changes & Changes \\
    \midrule
    BigVul & 13 (21\%) & 29 (48\%) & 19 (31\%) \\
    Devign & 27 (54\%) & 19 (38\%) & 4 (8\%) \\
    DiverseVul & 1 (3\%) & 23 (66\%) & 11 (31\%) \\
    \bottomrule
  \end{tabular}
\end{table}

\textbf{Unrelated Changes.} The third reason for label inaccuracy are other unrelated changes to functions in vulnerability patch commits. These include stylistic changes, e.g. removing whitespace or adding comments. According to our study, 8\%-31\% of falsely labeled functions can be attributed to such unrelated changes. 

Croft et al. \cite{data_quality_for_se_datasets} also investigate reasons for label inaccuracy and list irrelevant code changes (our structural changes), inaccurate fix identification (our patch commit identification), and clean-up changes (our unrelated changes). Based on our evidence, we can confirm these findings.

\subsection*{RQ.1-b\quad Context-Dependent Vulnerability}
\label{sec:results:rq_1:b}

The main goal of our empirical study was to find out what proportion of the functions labeled as vulnerable in the top-most widely-used datasets actually can be detected without considering additional context. In other words, what proportion of vulnerable functions would not be vulnerable if the appropriate external context did not exist?

\textbf{Results.} Each of the actual vulnerable functions that resulted from the first step of our empirical study (RQ.1 (a)) was assigned one of two labels: Context-independent or context-dependent. \autoref{fig:results:rq_1} show the results of this second labeling round. To our surprise, all 151 vulnerable functions in our study required additional context to be accurately identified (context-dependent). Not a single function could be detected without considering any additional context beyond the function itself (context-independent)

\result{\textbf{Context Dependence (RQ1-b).} All 151 vulnerable functions in our empirical study required additional context to be identified. Without this additional context, it was impossible to determine whether these functions actually contain security vulnerabilities. }

Based on the explanations we generated for each of the functions, we performed an in-depth analysis and identified the most prevalent types of context dependence in our sample. The results are shown in \autoref{table:reasons:context-dependence}.

\begin{table}[t]
%\scriptsize
\centering
  \caption{\textbf{Reasons for Context-dependence.} The different types of context dependence we observed and their prevalence in the three datasets.}
  \label{table:reasons:context-dependence}
  \begin{tabular}{@{}l@{ }|@{ }c@{\quad}c@{\quad}c@{\quad}c@{\quad}c@{}}
    \toprule
     & Function & External & Type & & Execution \\
    Dataset & Argument & Function & Declaration & Globals & Environment \\
    \midrule
    BigVul & 16 (41\%) & 19 (49\%)& 1 (2\%)& 3 (8\%) & 0 (0\%) \\
    Devign & 22 (47\%) & 20 (43\%)& 0 (0\%)& 5 (10\%) & 0 (0\%) \\
    DiverseVul & 26 (40\%) & 34 (52\%)& 2 (3\%)& 1 (2\%) & 2 (3\%) \\
    \bottomrule
  \end{tabular}
\end{table}

\textbf{Dependence on External Functions.} The first and most prevalent type of dependence we identified is dependence on external functions. Consider the example in \autoref{fig:dependence_types:a}. The heap-based buffer overflow in lines 20-21 of this function from the BigVul dataset depends on the external function \texttt{cfg80211\_find\_vendor\_ie}. Without knowing this external function, we do not know what values \texttt{vs\_ie} can have, and consequently, we do not know whether the buffer overflow can ever be triggered. In our empirical study, 43\%-52\% of context-dependent vulnerabilities can be attributed to dependence on external functions.

\textbf{Dependence on Function Arguments.} The second type of dependence we identified is dependence on function arguments. Consider the example in \autoref{fig:dependence_types:c}. The null-pointer-dereference vulnerability in line 8 of this function from the BigVul dataset depends on the function argument \texttt{sprinc}. However, without knowing the context in which this function can be called, we do not know whether \texttt{sprinc} can ever be \texttt{NULL}. For example, it could be properly validated before passing it to the function in all cases where this function is actually called. In that case, there would be no vulnerability. In our empirical study, 40\%-47\% of context-dependent vulnerabilities can be attributed to dependence on function arguments.

\textbf{Dependence on Type Declarations.} The third type of dependence we identified is dependence on type declarations. Consider the example in \autoref{fig:dependence_types:b}. The integer overflow vulnerability in line 20 depends on the type declaration of \texttt{PP\_Flash\_MenuItem}. Only if \texttt{sizeof(PP\_Flash\_MenuItem) * menu->count} exceeds the
range of \texttt{menu->items}, the operation overflows. In our empirical study, 0-3\% of context-dependent vulnerabilities can be attributed to dependence on type declarations.

\textbf{Dependence on Globals.} The fourth type of dependence we identified is dependence on globals, such as macros and global variables. Consider the example in \autoref{fig:dependence_types:d}. Line 7 of this function from the Devign dataset is the cause of multiple buffer overflows in the JasPer repository (CVE-2014-8158). If \texttt{HAVE\_VLA} is not defined and \texttt{QMFB\_JOINBUFSIZE} is smaller than \texttt{bufsize} but still used without successful dynamic allocation, any attempt to use \texttt{joinbuf} for storing data will result in writing beyond its allocated size (\texttt{QMFB\_JOINBUFSIZE}). In our empirical study, 2-10\% of context-dependent vulnerabilities can be attributed to dependence on globals.

\begin{figure}[t]
    \centering
    \begin{minipage}{\linewidth}
        \lstset{belowskip=0pt}
\begin{lstlisting}[lastline=19, frame=t, language=C, linewidth=0.9\linewidth]
static int mwifiex_update_vs_ie(const u8 *ies, int ies_len,
                                struct mwifiex_ie **ie_ptr, u16 mask,
                                unsigned int oui, u8 oui_type)
{
    struct ieee_types_header *vs_ie;
    struct mwifiex_ie *ie = *ie_ptr;
    const u8 *vendor_ie;

    vendor_ie = cfg80211_find_vendor_ie(oui, oui_type, ies, ies_len);
    if (vendor_ie) {
        if (!*ie_ptr) {
            *ie_ptr = kzalloc(sizeof(struct mwifiex_ie),
                              GFP_KERNEL);
            if (!*ie_ptr)
                    return -ENOMEM;
            ie = *ie_ptr;
         }

         vs_ie = (struct ieee_types_header *)vendor_ie;
\end{lstlisting}
        \lstset{aboveskip=0pt}
\begin{lstlisting}[firstline=20,lastline=21, firstnumber=20, language=C, backgroundcolor=\color{orange!30}, linewidth=0.9\linewidth]
         memcpy(ie->ie_buffer + le16_to_cpu(ie->ie_length),
                vs_ie, vs_ie->len + 2);
\end{lstlisting}
        \lstset{aboveskip=0pt,belowskip=5pt}
\begin{lstlisting}[firstline=22, firstnumber=22, frame=b, language=C, linewidth=0.9\linewidth]
         le16_unaligned_add_cpu(&ie->ie_length, vs_ie->len + 2);
         ie->mgmt_subtype_mask = cpu_to_le16(mask);
         ie->ie_index = cpu_to_le16(MWIFIEX_AUTO_IDX_MASK);
    }

    *ie_ptr = ie;
    return 0;
}
\end{lstlisting}
    \end{minipage}
    \caption{\textbf{Dependence on external functions:} The heap-based buffer overflow (CVE-2019-14816) in lines 20-21 of this function from the BigVul dataset depends on the external function \texttt{cfg80211\_find\_vendor\_ie}. Without knowing this external function, we do not know what values \texttt{vs\_ie} can have, and consequently, we do not know whether the buffer overflow can ever be triggered.}
    \label{fig:dependence_types:a}
\end{figure}

\begin{figure}[t]
    \centering
    \begin{minipage}{\linewidth}
        \lstset{belowskip=0pt}
\begin{lstlisting}[lastline=7, frame=t, language=C, linewidth=0.9\linewidth]
setup_server_realm(krb5_principal sprinc)
{
    krb5_error_code     kret;
    kdc_realm_t         *newrealm;
    
    kret = 0;
    if (kdc_numrealms > 1) {
\end{lstlisting}
        \lstset{aboveskip=0pt}
\begin{lstlisting}[firstline=8,lastline=8, firstnumber=8, language=C, backgroundcolor=\color{orange!30}, linewidth=0.9\linewidth]
        if (!(newrealm = find_realm_data(sprinc->realm.data, (krb5_ui_4) sprinc->realm.length)))
\end{lstlisting}
        \lstset{aboveskip=0pt,belowskip=5pt}
\begin{lstlisting}[firstline=9, firstnumber=9, frame=b, language=C, linewidth=0.9\linewidth]
            kret = ENOENT;
        else
            kdc_active_realm = newrealm;
    }
    else
        kdc_active_realm = kdc_realmlist[0];
    return(kret);
}
\end{lstlisting}
    \end{minipage}
    \caption{\textbf{Dependence on function argument:} The null pointer dereference (CVE-2013-6800) in line 8 of this function from the BigVul dataset depends on the function argument \texttt{sprinc}. Only if \texttt{sprinc=NULL} it crashes with a null pointer dereference in line 8.}
    \label{fig:dependence_types:c}
\end{figure}

\textbf{Dependence on the Execution Environment.} The fifth type of dependence we identified is dependence on the execution environment. For example, a vulnerability may depend on the fulfillment of specific conditions in the system of the user (e.g., the presence of a file in a directory) outside of the code. In our empirical study, 0-3\% of context-dependent vulnerabilities can be attributed to dependence on the execution environment.

\begin{figure}[t]
    \centering
    \begin{minipage}{\linewidth}
        \lstset{belowskip=0pt}
\begin{lstlisting}[lastline=19, frame=t, language=C, linewidth=0.9\linewidth]
PP_Flash_Menu* ReadMenu(int depth, const IPC::Message* m, PickleIterator* iter) {
  if (depth > kMaxMenuDepth)
    return NULL;
  ++depth;

  PP_Flash_Menu* menu = new PP_Flash_Menu;
  menu->items = NULL;

  if (!m->ReadUInt32(iter, &menu->count)) {
    FreeMenu(menu);
    return NULL;
  }

  if (menu->count == 0)
    return menu;
 
  menu->items = new PP_Flash_MenuItem[menu->count];
  memset(menu->items, 0, sizeof(PP_Flash_MenuItem) * menu->count);
  for (uint32_t i = 0; i < menu->count; ++i) {
\end{lstlisting}
        \lstset{aboveskip=0pt}
\begin{lstlisting}[firstline=20,lastline=20, firstnumber=20, language=C, backgroundcolor=\color{orange!30}, linewidth=0.9\linewidth]
    if (!ReadMenuItem(depth, m, iter, menu->items + i)) {
\end{lstlisting}
        \lstset{aboveskip=0pt,belowskip=5pt}
\begin{lstlisting}[firstline=21, firstnumber=21, frame=b, language=C, linewidth=0.9\linewidth]
      FreeMenu(menu);
      return NULL;
    }
  }
  return menu;
}
\end{lstlisting}
    \end{minipage}
    \caption{\textbf{Dependence on type declaration:} The integer overflow (CVE-2013-0892) in line 20 of this function from the BigVul dataset depends on \texttt{PP\_Flash\_MenuItem}, \texttt{menu->items} and \texttt{menu->count}. Only if \texttt{sizeof(PP\_Flash\_MenuItem) * menu->count} exceeds the range of \texttt{menu->items}, the operation overflows. }
    \label{fig:dependence_types:b}
\end{figure}

\begin{figure}[t]
    \centering
    \noindent\begin{minipage}{\linewidth}
        \lstset{belowskip=0pt}
\begin{lstlisting}[lastline=6, frame=t, language=C, linewidth=0.9\linewidth]
void jpc_qmfb_join_col(jpc_fix_t *a, int numrows, int stride,
  int parity)
{

        int bufsize = JPC_CEILDIVPOW2(numrows, 1);
#if !defined(HAVE_VLA)
\end{lstlisting}
        \lstset{aboveskip=0pt}
\begin{lstlisting}[firstline=7,lastline=7, firstnumber=7, language=C, backgroundcolor=\color{orange!30}, linewidth=0.9\linewidth]
        jpc_fix_t joinbuf[QMFB_JOINBUFSIZE];
\end{lstlisting}
        \lstset{aboveskip=0pt,belowskip=5pt}
\begin{lstlisting}[firstline=8, firstnumber=8, frame=b, language=C, linewidth=0.9\linewidth]
#else
        jpc_fix_t joinbuf[bufsize];
#endif
        jpc_fix_t *buf = joinbuf;

        // [...]

}
\end{lstlisting}
    \end{minipage}
    \caption{\textbf{Dependence on globals:} Line 7 of this function is the cause of multiple buffer overflows in the JasPer repository (CVE-2014-8158). Whether line 7 is actually compiled depends on whether the macro \texttt{HAVE\_VLA} is defined.}
    \label{fig:dependence_types:d}
\end{figure}

\subsection*{RQ.1-c\quad Context-dependent Security}
\label{sec:rq_3_results}
Now we know that the vulnerability of all 151 \texttt{vulnerable} functions in our empirical study could not be decided without considering additional context, but what about \texttt{secure} functions? What proportion of functions labeled as \texttt{secure}---because they are not \emph{known} to cause a vulnerability in the program---could potentially cause a vulnerability if an appropriate external context existed?

\textbf{Methodology.}
From each of the three datasets (BigVul, Devign, and DiverseVul), we randomly selected 30 samples that were labeled as \texttt{secure} and checked if we could come up with a reasonable context for that function that would cause a vulnerability in the program. We note that we did not attempt to verify whether a function that was labeled as \texttt{secure} was \emph{actually} not causing a vulnerability in the program. There can only be certainty about the \emph{presence} of vulnerabilities (e.g., via a triggering input). In practice, there is no certainty about their \emph{absence}. However, we publish the reproducible script and the sampled functions as part of our artifact for full transparency. 

To find out what proportion of \texttt{secure} functions could be vulnerable if an appropriate external context would exist, we tried to construct an artificial vulnerable context for each of them. Consider the example in \autoref{fig:crafted_context}. In this artificially crafted context, the function \texttt{HTTP\_Clone} (marked in green) contains a null pointer dereference vulnerability. However, it is labeled as \texttt{secure} in the DiverseVul dataset because in its original context\footnote{\url{https://github.com/varnishcache/varnish-cache/commit/c5fd097e}} it is not known to contain a security vulnerability. Just looking at that function itself, without any external context, we cannot decide whether this function does not actually cause a vulnerability in the program. Similar to this example, we manually tried to construct a vulnerable context for all 90 functions in our sample. For full transparency, we include the generated vulnerable settings (as explanations in natural language) in our artifact. 

\textbf{Results.} For 82 out of the 90 functions that were labeled as secure in the original datasets, we were able to construct a context in which they contain a security vulnerability. 
\result{\textbf{Secure Functions (RQ1-c).} 82 out of 90 secure functions contain context-dependent vulnerabilities, i.e., they would make the program vulnerable if the appropriate external context existed. In isolation, it is not possible to determine whether these functions are actually secure. }

\subsection*{RQ.1\quad Result Summary}

Within our sample of functions, about half of those \emph{labeled as vulnerable} did not actually contain the security vulnerabilities they were claimed to have. All of those that do are context-dependent vulnerabilities, i.e., these functions are vulnerable only \emph{because} an appropriate context exists under which the function is vulnerable. Similarly, the majority of secure functions contain context-dependent vulnerabilities that would make that function vulnerable if an appropriate context existed.
As there is evidently insufficient information in the base units---that are used for training, validation, and testing in these datasets---to decide their vulnerability without further context, we conclude that ML4VD cannot be soundly evaluated as a classic binary classification problem on the function-level. 
\result{\textbf{Problem statement (RQ.1)}. ML4VD techniques cannot be soundly evaluated as a classic binary classification problem on intra-procedural base units, such as functions.}

\begin{figure}[t]
\centering
    \begin{minipage}{\linewidth}
        \lstset{belowskip=0pt}
\begin{lstlisting}[lastline=11, frame=t, language=C, linewidth=0.9\linewidth]
#include <stdio.h>
#include <stdlib.h>

struct http {
    int vsl;
    int ws;
};

void HTTP_Dup(struct http *to, const struct http *fm) {
    return;
}
\end{lstlisting}
        \lstset{aboveskip=0pt}
\begin{lstlisting}[firstline=12,lastline=17, firstnumber=12, language=C, backgroundcolor=\color{green!30}, linewidth=0.9\linewidth]
void HTTP_Clone(struct http *to, const struct http *fm) {
    HTTP_Dup(to, fm);
    to->vsl = fm->vsl;
    to->ws = fm->ws;
}
\end{lstlisting}
        \lstset{aboveskip=0pt,belowskip=5pt}
\begin{lstlisting}[firstline=18, firstnumber=18, frame=b, language=C, linewidth=0.9\linewidth]
int main() {
    struct http *source = NULL;
    struct http destination;

    HTTP_Clone(&destination, source);

    return 0;
}
\end{lstlisting}
    \end{minipage}

    \caption{\textbf{Context-dependent security:} In this artificially crafted context, the function HTTP\_Clone contains a null pointer dereference vulnerability. However, it is labeled as 'secure' in the DiverseVul dataset because in its original context (see \url{https://github.com/varnishcache/varnish-cache/commit/c5fd097e}) it is not known to contain a security vulnerability. }
    \label{fig:crafted_context}
\end{figure}

\subsection*{RQ.2\quad Classifier Performance on Spurious Features}
\label{sec:rq_4_results}

Since it is impossible for most functions to decide without further context whether they contain a vulnerability, there is currently no evidence that ML4VD techniques are actually capable of identifying security vulnerabilities in functions. But why do ML4VD papers still report high scores when evaluating their techniques using function-level datasets? What happens if we hide the code and only expose some features of the code, such as word counts?

\textbf{Results.} \autoref{table:spurious-correlations} shows the results for RQ.2. The Gradient Boosting Classifier achieved a f1-score of 86\% on the evaluation subset of BigVul, a f1-score of 62.2\% and an accuracy of 63.2\% on the evaluation subset of Devign, and a f1-score of 11.1\% on the evaluation subset of DiverseVul. For Devign and BigVul, the observed performances are only 4.4\% (Devign, accuracy), 5.7\% (Devign, f1-score), and 10\% (BigVul) lower than the performances of recently published state-of-the-art techniques. Surprisingly, for an effective vulnerability detection model, the whole process of training and evaluation only took 30 minutes on a MacBook Pro, which is extremely fast compared to training times of state-of-the-art ML-based techniques. State-of-the-art ML-based techniques require expensive Hardware (GPUs), weeks of computing time to pre-train, and at least multiple hours of computing time to finetune. Additionally, the Gradient Boosting Classifier was only trained on the training subset of the two datasets without any pre-training. These results show that it is possible to achieve high performance on the most popular function-level datasets while completely disregarding the structure and semantics of the code. 

\begin{table}[t]
%\scriptsize
\centering\small
  \caption{\textbf{Spurious Correlations.} The performance of a simple model (Gradient Boosting Classifier) trained on word counts only (no information on code structure and/or semantics available during training).}
  \label{table:spurious-correlations}
  \begin{tabular}{@{}l@{ }|@{ }c@{\quad}c@{}}
    \toprule
    Dataset & State-of-the-art & Without Structure \\
    \midrule
    BigVul (f1-score) & 0.96 & 0.86 \\
    Devign (f1-score) & 0.68 & 0.62 \\
    DiverseVul (f1-score) & 0.48 & 0.11 \\
    Devign (accuracy) & 0.68 & 0.63 \\
    \bottomrule
  \end{tabular}\vspace{-0.2cm}
\end{table}

For the DiverseVul dataset, the Gradient Boosting Classifier was not able to achieve a comparable performance to the state-of-the-art techniques. This suggests that the dataset may contain fewer exploitable spurious features. However, this is not definitive proof. What the result does show is that state-of-the-art performance cannot be reached using a Gradient Boosting Classifier based solely on word count features. 

\result{\textbf{Spurious Features  (RQ2).} Using word counts only, we were able to achieve 62.2\% f1-score on the Devign dataset and 86\% f1-score on the BigVul dataset with a simple Gradient Boosting Classifier. These results show that the top-most widely-used function-level datasets can be exploited to achieve high scores without actually detecting security vulnerabilities.}

Our results provide an alternative explanation for the results reported in the literature. While we did not prove that state-of-the-art ML4VD techniques actually achieve their high scores by relying on spuriously correlated features, we have shown that it is possible to exploit these datasets to achieve high scores without actually detecting security vulnerabilities. 

\section{Threats to the Validity}
\label{sec:threats_to_validity}

As for any empirical study, there are various threats to the validity of our results and conclusions.

\subsection{Internal validity} 

\textbf{Selection Bias.} The random sampling of 100 functions from each dataset (BigVul, Devign, and DiverseVul) for our empirical study could introduce selection bias. Although random sampling aims to create a representative subset, it is possible that our sample may not fully capture the diversity and characteristics of the entire dataset. To mitigate this, we ensured that our sampling method was strictly random and publish the reproducible script and the sampled functions as part of our GitHub repository: \url{https://github.com/niklasrisse/TopScoreWrongExam}.

\textbf{Manual Labeling Errors.} Labeling functions as vulnerable or secure and distinguishing between context-dependent and context-independent vulnerabilities involves subjective judgment. To address this, we employed a cross-labeling process where all 300 functions were independently labeled by two Software Security researchers, achieving substantial agreement (82\%) for labeling functions as vulnerable or secure and almost perfect agreement (98\%) for distinguishing between context-dependent and context-independent vulnerabilities. Discrepancies were resolved through discussion, but some human error may still be present. For full transparency, we publish all labels and explanations as part of our artifact.

\subsection{External validity}

\textbf{Datasets}. Our study focuses on three specific datasets (BigVul, Devign, and DiverseVul), which are widely used in the ML4VD community (cf. Fig.~\ref{fig:datasets}). However, the findings may not generalize to other datasets or real-world software systems. Future studies should replicate our methodology on additional datasets and real-world codebases to validate our conclusions.

\textbf{Programming languages}. The datasets analyzed primarily contain C code. Our findings might not generalize to other programming languages with different syntactic and semantic properties. Future research should include datasets from various programming languages to evaluate the broader applicability of our results.

\textbf{Problem Statements}. Our study focuses on the most prevalent problem statement of ML4VD as function-level binary classification problem (cf. \S\ref{sec:literature_survey} and Fig.~\ref{fig:survey}) where the classifier makes the decision exclusively based on information within the given function. However, there exist other definitions of ML4VD (e.g., at other granularities, such as line- or file-level, and even some that explicitly consider context, such as vulnerability-specific inter-procedural slicing \cite{fse_slicing}). While we discuss our perspective for other definitions of ML4VD in \autoref{sec:discussion}, future studies should carry out similar investigations to validate how reasonable these problem statements are indeed.

\subsection{Construct Validity}

\textbf{Spurious Correlations.} Our results suggest that ML4VD techniques might achieve high scores by exploiting spurious correlations rather than genuinely detecting vulnerabilities. While we demonstrated this using word counts, further research should investigate other features that might be \emph{spuriously} correlated with the \texttt{vulnerable} label. We discuss some related work in \autoref{sec:related_work}.

\section{Discussion and Future Work}
\label{sec:discussion}

The goal of empirical evaluation in ML4VD is to assess the capabilities of specific techniques by putting them to the test on real-world data. However, if the evaluation is based on incorrect assumptions, the results become meaningless. Our findings demonstrate that this is precisely the issue with ML4VD. Techniques can achieve ‘top scores’ by exploiting spurious features, even without possessing the capabilities the test aims to measure. In our specific case, the only plausible explanation for the good performance of ML4VD techniques are spurious correlations.

\textbf{Implications for general ML.} Are we truly measuring the effectiveness of ML techniques in solving the specific tasks we expect them to solve? The issue of spurious correlations is not limited to ML4VD but extends to other areas of machine learning, where it has been explored in depth by several studies \cite{spurious_ml_1, spurious_ml_2}. It is crucial to continue the development of methods to identify and measure these correlations during the benchmarking of ML techniques to ensure that we are genuinely evaluating their performance in solving the tasks they are intended to address. 

\textbf{Implications for Program Analysis.} The prevalence of the context-dependency problem suggests that an effective function-level program analysis, such as an \emph{intra}-procedural static analysis, must consider a reasonable approximation of the space of valid external states during the average function call, e.g., encoded as a function precondition, function summary, or type system. While the context-dependency check can ultimately be delegated to the user in practice, given our results, we suggest that particular care is taken when evaluating such tools at the function-level in the absence of users. In addition to function-level program analysis, there are \emph{inter}-procedural, or whole-program analyses, which consider interactions between multiple procedures, and thus implicitly resolve the context-dependency problem during the search for security vulnerabilities.

\textbf{Abstention.} What could be possible ways forward for ML4VD? A simple possible solution to the context-dependency problem of function-level vulnerability detection could be to cast it as binary classification \emph{with abstention}. Given a function, an ML4VD technique could either decide the vulnerability of the function or abstain from this decision. However, the obvious disadvantage of this approach is that only context-independent security vulnerabilities could actually be detected, which appear to be very rare (cf. RQ.1-b). The context-dependency labels generated by our empirical study could be used as a starting point to explore this direction.

\textbf{Other Base Units.} Instead of deciding whether a \emph{function} is \texttt{vulnerable}, we could decide whether other types of base units are \texttt{vulnerable}, such as lines, statements, files, modules, commits, or vulnerability-specific inter-procedural slices (cf. Fig.~\ref{fig:survey}). However, for smaller base units, such as line-, statement-, or commit-level, there is no reason to believe that the context-dependency problem is resolved (subject to further study) while for larger base units, such as files or modules, the context-dependency problem may be much less pronounced (subject to further study), although the \texttt{vulnerability} label at that such coarse granularity may be less actionable for the security researcher.
For alternative granularities, future work shall determine if they offer a true advantage or merely shift the problem to a different level of abstraction.

The vulnerability-specific inter-procedural slice \cite{fse_slicing} (at least conceptually) includes all required context to trigger a vulnerability of a specific type at a specific code location, and thus resolves the context-dependency problem. However, it also somewhat delegates the detection problem because it requires, as slicing criterion, a (vulnerability-specific) statement-of-interest where a vulnerability may be triggered. In general, we believe that a combination of static analysis and machine learning \cite{ml_and_static_analysis} offers a promising avenue for future research in vulnerability detection.

\textbf{Context-conditional Classification.} 
We could define ML4VD as a base-unit-level classification problem \emph{given}, e.g., the entire software repository as context.
Due to the hierarchical nature of software, determining the vulnerability of a given base unit (e.g., function) may require considering a large scope of context, such as the complete state of a repository at a given commit ID and all other dependencies that might exist outside of the code (e.g. system of the user, dependencies of the repository, etc.). Since the Devign, BigVul, and DiverseVul benchmarks all include patch commit IDs, the complete context for a given function could be reconstructed, so that they could be used to evaluate ML4VD based on this alternative problem statement. Future research is needed to find optimal ways to include this context into state-of-the-art techniques.

However, adding repository-level context alone may not resolve the benchmarking problem. Evidently, ML4VD techniques that \emph{disregard} this context still appear to perform well (likely due to spurious correlations; RQ.2). Addressing this challenge will require ensuring that the evaluation metrics and benchmarks \emph{actually} capture the vulnerability detection capabilities of the techniques that are tested.

\textbf{Overcoming Classification.} Even using different types of base units and providing the entire repository as context, ML4VD would still be a binary classification task: Given a unit, decide whether the unit is \texttt{vulnerable} or \texttt{secure}.
Alternatively, an ML4VD technique could \emph{provide the condition under which a given function would actually be vulnerable}, e.g., by generating an example context: Given a function, an ML4VD technique may generate a complete executable program which is vulnerable to attack due to the code in that function. This generated context could then be compared with the actual context to see whether the vulnerability exists in the real context. 
A much stronger approach could be the (ML-assisted) generation of a proof of vulnerability in form of a test case, which demonstrates the exploitation of the vulnerability in the given context.

In conclusion, addressing the issues of context dependency and spurious correlations is critical for the advancement of ML4VD and other ML applications. By exploring alternative methodologies and improving our evaluation frameworks, we can ensure more robust and reliable assessments, ultimately leading to more secure and effective solutions.

\section{Data Availability}
\label{sec:data-availability}
To ensure the reproducibility of our results and to provide transparency in our research, we have made all related scripts and data publicly available. All resources can be accessed as part of our GitHub repository, which is available at \url{https://github.com/niklasrisse/TopScoreWrongExam}.

\section*{Acknowledgments}
Funded by the Deutsche Forschungsgemeinschaft (DFG, German Research Foundation) under Germany's Excellence Strategy - EXC 2092 CASA - 390781972.
This research
is also partially funded by the European Union. Views
and opinions expressed are however those of the author(s)
only and do not necessarily reflect those of the European
Union or the European Research Council Executive Agency.
Neither the European Union nor the granting authority can
be held responsible for them. This work is supported by
ERC grant (Project AT\_SCALE, 101179366).

\bibliographystyle{IEEEtran}
\bibliography{main}

\end{document}